\documentclass[10pt,journal,compsoc]{IEEEtran}

\usepackage{amsmath}
\usepackage{cite}
\usepackage{graphicx}
\usepackage{footnote}
\usepackage{bbding}
\usepackage{pifont}
\usepackage{wasysym}
\usepackage{amssymb}
\usepackage{multirow}
\usepackage[table,xcdraw]{xcolor}
\usepackage{diagbox}
\usepackage[flushleft]{threeparttable}
\makesavenoteenv{tabular}
\usepackage{tabu}
\usepackage{longtable}
\usepackage{cite}
\usepackage{epstopdf}
\usepackage{color}
\usepackage{enumitem}
\usepackage{booktabs}
\usepackage[noend]{algpseudocode}
\usepackage[ruled]{algorithm2e}
\usepackage{subfig}

\setlist{parsep=0pt,listparindent=\parindent}

\ifCLASSINFOpdf
\else
\fi

\usepackage{ragged2e}

\begin{document}
%
\title{Fast Gradient Attack on Network Embedding}
%
%
%
%

\author{Jinyin~Chen, Yangyang~Wu, Xuanheng~Xu, Yixian~Chen, Haibin~Zheng, and~Qi~Xuan~\IEEEmembership{IEEE Member}%
\IEEEcompsocitemizethanks{
\IEEEcompsocthanksitem J. Chen, Y. Wu, X. Xu, Y. Chen, H Zheng, and Q. Xuan are with the College of Information Engineering, Zhejiang University
of Technology, Hangzhou 310023, China. E-mail: \{chenjinyin, 2111603080, 2111603112, 201407760108, 201303080231, xuanqi\}@zjut.edu.cn.\protect\\
\IEEEcompsocthanksitem This article has been submitted on August 18th, 2018.\protect\\
\IEEEcompsocthanksitem Corresponding author: Qi Xuan}
}

\IEEEtitleabstractindextext{%
\begin{abstract}
\justifying
Network embedding maps a network into a low-dimensional Euclidean space, and thus facilitate many network analysis tasks, such as node classification, link prediction and community detection etc, by utilizing machine learning methods. In social networks, we may pay special attention to user privacy, and would like to prevent some target nodes from being identified by such network analysis methods in certain cases. Inspired by successful adversarial attack on deep learning models, we propose a framework to generate adversarial networks based on the gradient information in Graph Convolutional Network (GCN). In particular, we extract the gradient of pairwise nodes based on the adversarial network, and select the pair of nodes with maximum absolute gradient to realize the Fast Gradient Attack (FGA) and update the adversarial network. This process is implemented iteratively and terminated until certain condition is satisfied, i.e., the number of modified links reaches certain predefined value. Comprehensive attacks, including unlimited attack, direct attack and indirect attack, are performed on six well-known network embedding methods. The experiments on real-world networks suggest that our proposed FGA behaves better than some baseline methods, i.e., the network embedding can be easily disturbed using FGA by only rewiring few links, achieving state-of-the-art attack performance.

\justifying
\end{abstract}

}

\maketitle

\IEEEdisplaynontitleabstractindextext

%
\IEEEpeerreviewmaketitle

\ifCLASSOPTIONcompsoc
\IEEEraisesectionheading{\section{Introduction}\label{sec:introduction}}
\else
\section{Introduction}
\label{sec:introduction}
\fi

%
%
%
%
\IEEEPARstart{O}{ur} lives are surrounded by various networks, such as social networks, communication networks, biological networks, traffic networks and so on. Network embedding~\cite{cai2018comprehensive,wang2017knowledge,Choi2017GRAM}, used to learn low-dimensional representations for nodes or links in the network, is capable to benefit a wide range of real-world applications such as link prediction~\cite{perozzi2014deepwalk,Wang2017Signed}, node classification~\cite{Tang2015PTE,Wang2016Linked}, community detection~\cite{Tian2014Learning,allab2017semi}, social network analysis~\cite{liu2016aligning} etc. The embedding methods will directly determine the performances of downstream applications, and thus have been receiving more and more attentions in the past decades.

Some earlier works, including IsoMAP~\cite{Tenenbaum2000A}, local linear embedding~\cite{Roweis2000Nonlinear} and Laplacian eigenmap~\cite{belkin2002laplacian}, tried to embed the network by decomposing the similarity matrix. In the last few years, on the other hand, more and more studies focused on embedding the network into a low-dimensional vector space. For instance, since Mikolov et al.~\cite{Mikolov2013Efficient} proposed the word2vec model, the skip-gram mechanism is widely adopted in many network embedding methods, such as DeepWalk~\cite{perozzi2014deepwalk}, LINE~\cite{tang2015line} and node2vec~\cite{grover2016node2vec} etc. DeepWalk~\cite{perozzi2014deepwalk} was the first model to learn language from a network, which adopts random walk to sample a sequence of nodes for each node, and then treats these sequences as sentences by the skip-gram mechanism. LINE~\cite{tang2015line} can be considered as a special case of DeepWalk, with the window size of contexts set to 1. Node2vec~\cite{grover2016node2vec} is an extension of DeepWalk, which is more flexible when generating the context of a node. The generated contexts of nodes are also treated as text in a language model to learn the embeddings by the skip-gram mechanism. Another embedding method, namely GraRep, is proposed by Cao et al.~\cite{cao2015grarep}, which preserves node proximities by constructing different k-step probability transition matrices.

Quite recently, a few deep embedding methods were proposed, which are generally based on generative adversarial networks (GAN)~\cite{Goodfellow2014Generative} and graph convolutional network (GCN)~\cite{kipf2016semi}. For example, Wang et al.~\cite{Wang2017GraphGAN} proposed GraphGAN as a novel network representation learning framework, which unifies two classes of graph representation learning methodologies via adversarial training in a minimax game. Dai et al.~\cite{dai2017adversarial} proposed an adversarial network embedding method, which adopts the adversarial learning principle to regularize the representation learning. On the other hand, Kipf et al.~\cite{kipf2016semi} proposed the GCN as a basic graph convolution method for semi-supervised classification, which learns the hidden layer representations that encode both local graph structure and features of nodes. Moreover, Pham et al.~\cite{pham2017column} introduced Column Network (CLN) as a novel deep learning model for collective classification, inspired by the columnar organization of neocortex.

Although deep learning methods achieve great success in many real-world tasks, such as computer vision~\cite{Xuan2017Automatic}, natural language process~\cite{Sak2014Long} and so on, they are confronted with security problem~\cite{goodfellow2016deep}. The most typical one is adversarial attack~\cite{goodfellow2014explaining,moosavi2016deepfool,elsayed2018adversarial,carlini2017towards,kurakin2018adversarial,papernot2017practical,moosavi2017universal,biggio2014security,mei2015using}, i.e., in computer vision tasks, we can add a designed tiny perturbation into an original image to fool a CNN model, leading to the wrong classification of the image~\cite{goodfellow2014explaining}. The study of network analysis attacks, on the other hand, roots in the need for protecting the user privacy from, or understanding the robustness of, those state-of-the-art network analysis methods.

For instance, in community detection, Nagaraja~\cite{nagaraja2010impact} proposed the first community deception method by adding links to the nodes of high centrality. Inspired by modularity, Waniek et al.~\cite{waniek2018hiding} proposed a scalable heuristic method, namely Disconnect Internally, Connect Externally (DICE), which randomly deletes the links between the nodes in the target community, while adds the links between them and those of different communities. Besides, Fionda et al.~\cite{fionda2018community} proposed a novel community deception method based on the safeness which evaluates the hiding level of a target community in the output of a detection algorithm. In link prediction, Zheleva et al.~\cite{zheleva2008preserving} proposed a link re-identification attack to inferring sensitive links from the released data. Link perturbation is a common technique in early research that data publisher can randomly modify links on the original network to protect the sensitive links from being identified. Fard et al.~\cite{fard2012limiting} introduced a subgraph-wise perturbation in directed networks to randomize the destination of a link within subgraphs to protect sensitive links; they further proposed a neighborhood randomization mechanism to probabilistically randomize the destination of a link within a local neighborhood~\cite{fard2015neighborhood}.

More interestingly, Z{\"u}gner et al.~\cite{DBLP:conf/kdd/ZugnerAG18} focused on the node classification using GCN, and proposed the first adversarial attacks on networks, namely NETTACK, which generated adversarial network iteratively. In each iteration, it first selected candidate links and features based on their important data characteristics such as degree distribution and co-occurence of features; then, it defined two scoring functions to evaluate the change in the confidence value of the target node after modifying a link and feature in the candidate sets, respectively; after that, it used the link or feature of the highest score to update the adversarial network. However, this approach is limited to node classification task, with little discussion on the transferability of the attack. On the other hand, while network embedding methods are getting more and more popular in network analysis, their security problem is largely ignored. We argue that the security problem of network embedding is more crucial than that of a particular network analysis method in community detection or link prediction, because if an embedding method is attacked, all downstream applications based on the obtained embedding vectors could be affected correspondingly.


Inspired by~\cite{DBLP:conf/kdd/ZugnerAG18}, in this paper, we propose a new fast gradient attack (FGA) on network embedding. Specifically, we make the following contributions.
\begin{itemize}
\item First, we design an adversarial network generator, utilizing the iterative gradient information of pairwise nodes based on the trained GCN model to generate adversarial network so as to realize the FGA.
\item Second, we use FGA to attack not only GCN model but also several other network embedding methods, such as GraRep, DeepWalk, node2vec, LINE and GraphGAN, and propose various adversarial attack strategies to attack multiple network analysis methods.
\item Third, the experiments validate that our FGA method performs significantly better than several advanced attack methods such as NETTACK and DICE, in terms of higher Attack Success Rates (ASR) and smaller Average number of Modified Links (AML) to successfully attack a target node, on a number of real-world networks, achieving state-of-the-art attack effects.
\end{itemize}




The rest of paper is organized as follows. In Sec.~\ref{Method}, we introduce our FGA method in details and explain the white-box and black-box adversarial attacks. In Sec.~\ref{Exp}, we empirically evaluate FGA on multiple tasks, and compare the attack effects by utilizing FGA and other attack methods on several real-world networks. In Sec.~\ref{Conclusion}, we conclude the paper and highlight future research directions.

\section{Method\label{Method}}
In this section, we introduce the framework of FGA on network embedding, where we propose an adversarial network generator based the GCN model. For convenience, the definitions of symbols used in this paper are briefly summarized in TABLE~\ref{Definition}.

\begin{table}[!t]
\centering
\caption{The definitions of symbols.}
 \label{Definition}
 \resizebox{\linewidth}{!}{
\begin{tabular}{lr}
\hline \hline
Symbol & Definition \\ \hline
   $G=(V,E)$           & input original network with nodes $V$ and links $E$           \\
   $\bar{G}=(V,\bar{E},M)$     & perturbation network with nodes $V$, links $\bar{E}$ and weight $M$\\
   $\hat{G}=(V,\hat{E})$       & attacked network with nodes $V$ and updated links $\hat{E}$           \\
   $A$           & the adjacency matrix of original network $G$ \\
   $\tilde{A}$   & the adjacency matrix added self-connections           \\
   $I_N$         & identity matrix           \\
   $\tilde{D}$   & degree matrix of $\tilde{A}$ \\
   $Y'(A)$       & the output of the GCN model           \\
$f$ and $\sigma$ & the softmax function and Relu active function           \\
   $X$           & the matrix of node feature vectors           \\
   $\bar{A}$     & the convolved signal matrix  \\
   $W_i$         & the weight matrices of GCN model          \\
   $C$           & the number of feature vector dimensions in $X$           \\
   $H$           & the number of feature maps for hidden layer          \\
   $F$           & the categories set for nodes in the network           \\
   $L$           & the loss function of the GCN model           \\
   $V_L$         & the set of nodes with labels           \\
   $Y$           & the real label confidence list           \\
   $\eta$        & learning rate           \\
   $v_t$         & target node           \\
   $L_{t}$       & target loss function for node $v_t$           \\
   $g$           & link gradient matrix           \\
   $\hat{g}$     & link gradient network          \\
   $K$           & the number of modified links           \\
   $\hat{G}^h$ & the $h^{th}$ adversarial network           \\
   $\hat{A}^h$ & the $h^{th}$ adversarial adjacency matrix           \\
   $\hat{g}^h$ & the $h^{th}$ link gradient network    \\
   $\theta$      & the sign function           \\
   $n_t$         & the number of links of the target node in $G$ \\
   $b$           & the number of links we random disconnect\\
   $\gamma$      & perturbation size\\   \hline  \hline
\end{tabular}}
\end{table}

\subsection{Problem Definition}
We first give the definitions of network embedding and network embedding attack as follows.

\begin{itemize}
\item \textbf{Network embedding:} It learns a mapping $: v_i\to y_i\in R^d$ of nodes in network $G=(V,E)$ to features in a low-dimensional space, based on which the downstream methods can be designed to realize node classification or clustering tasks etc. Generally, the dimension of each node $d$ is much smaller than the number of nodes $|V|$.
\item \textbf{Network embedding attack:} Given the network $G=(V,E)$, network embedding attack selects some key links for target nodes to construct the perturbation network $\bar{G} = (V,\bar{E},M)$, where $M_{ij} \in \{-1,0,1\}$ indicates the modification strategy of $\bar{E}_{ij}\in \bar{E}$. Then, the links $\hat{E}$ in the attacked network $\hat{G}=(V,\hat{E})$ is defined as
\begin{equation}
\hat{E}_{ij}=E_{ij}+M_{ij}\bar{E}_{ij}.
\label{DEFINITION:2}
\end{equation}
\end{itemize}
In this attacked network, the target nodes can be well hidden, e.g., they will be misclassified with a relatively high probability in a node classification task.

\subsection{The Framework of FGA}
\begin{figure*}[!t]
  \centering
  \includegraphics[width=\linewidth]{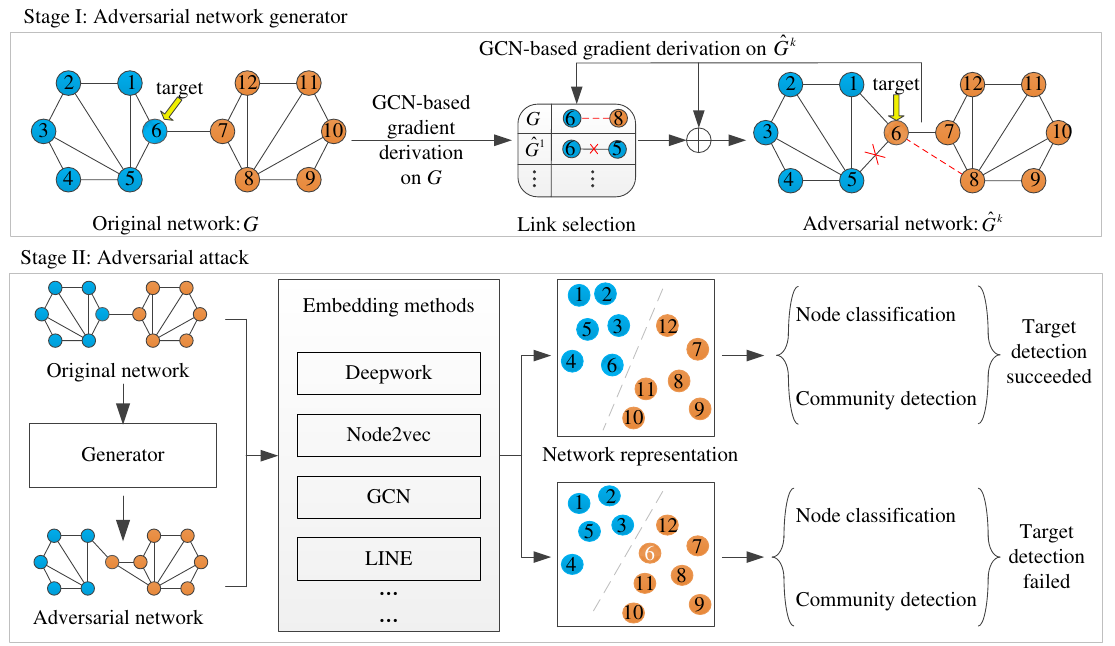}
  \caption{The framework of FGA on network embedding methods.}
 \label{framework}
\end{figure*}
Adversarial attack is launched by latent adversarial samples with minimized alternation from normal ones. In this study, the adversarial networks are elaborately designed to fool network embedding methods. When the original network is inputted, vectors of nodes are learned based on network embedding methods for a specific task, e.g., node classification, with satisfying performance. Then, we choose the target nodes, and generate adversarial networks to hide them, i.e., to make them misclassified. In other words, when the adversarial networks are inputted, the vectors of most nodes will keep the same as those in the original network, i.e., they will be correctly classified, while the target nodes will be unconsciously misclassified due to their significantly changed vectors. More remarkably, the adversarial networks are almost the same as the original one, e.g., only 1 to 6 links need to be rewired in a network of more than 5000 links in total. In particular, our FGA method consists of two stages: adversarial network generation and adversarial attack, as shown in Fig.~\ref{framework}.

\begin{itemize}
\item \textbf{Adversarial network generation:} Given a network, the adversarial network is generated based on the GCN gradient. First, we use the original network to train the GCN model. Then, for each target node, we design a target loss function, based on which we calculate the partial derivative, and further the gradient information, for each pair of nodes in the network. After that, we select the pair of nodes of the maximum \emph{absolute} gradient to update the adversarial network. Finally, we stop the process when a certain number of links are modified, and then output the final adversarial network.
\item \textbf{Adversarial attack:} We then use the generated adversarial network to protect the target node from the detection of GCN model. Since GCN has overwhelming generalization ability, the adversarial attack can be still effective for many other network embedding methods, i.e., the perturbation generated by GCN is universal and the attack thus has strong transferability.
\end{itemize}


\subsection{Adversarial Network Generator via GCN}{\label{generator}}
In this stage, we use the GCN model to generate adversarial networks, described as follows.

\subsubsection{GCN model}
We consider a two-layer GCN model for node classification in a network with the adjacency matrix $A$. $\tilde{A}=A+I_N$ is the adjacency matrix of the undirected network $G$ with the added self-connections, where $I_N$ is the identity matrix. $\tilde{D}_{ii}=\sum_j\tilde{A}_{ij}$ is the degree matrix of $\tilde{A}$.

We consider the GCN model with a single hidden layer, and its forward model takes the simple form:
\begin{equation}
Y'(A)=f(\bar{A}\sigma(\bar{A}XW_0)W_1),
\label{equ:forward}
\end{equation}
where $X$ is a matrix of node feature vectors, $\bar{A}=\tilde{D}^{-\frac{1}{2}}\tilde{A}\tilde{D}^{-\frac{1}{2}}=\tilde{D}^{-\frac{1}{2}}(A+I_N)\tilde{D}^{-\frac{1}{2}}$, $W_0\in R^{C\times H}$ and $W_1\in R^{H\times |F|}$ are the input-to-hidden and hidden-to-output weight matrices, respectively, with the hidden layer of $H$ feature maps; $f$ and $\sigma$ are the softmax function and Relu active function, respectively. Here, the softmax activation function is applied row-wise.

For node classification, we evaluate the cross-entropy error over all training examples:
\begin{equation}
L=-\sum_{l=1}^{|V_L|}\sum_{k=1}^{|F|}Y_{lk}\ln(Y'_{lk}(A)),
\label{equ:loss1}
\end{equation}
where $V_L$ is the set of nodes with labels, $F=[\tau_1,\cdots,\tau_{|F|}]$ is the category set for the nodes in the network, $|F|$ denotes the number of categories, $Y$ is the real label matrix with $Y_{lk}=1$ if node $v_l$ belongs to category $\tau_k$ and $Y_{lk}=0$ otherwise, and $Y'(A)$ is the output of the model calculated by Eq.~(\ref{equ:forward}). In the $m$-th iteration step, the weights $W_i$, $i\in\{0,1\}$, of the neural network model are trained using gradient descent, with the update role
\begin{equation}
W_i^{m+1}=W_i^m-\eta\frac{\partial L}{\partial W_i^m},
\label{equ:partial}
\end{equation}
where $\eta$ is the learning rate.

\subsubsection{Link gradient based on GCN}
According to Eq.~(\ref{equ:partial}), the weight matrices are updated based on the gradient information in the GCN model, so that the model is continuously optimized and the node classification performance is steadily improved. As we can see in Eq.~(\ref{equ:forward}) and Eq.~(\ref{equ:loss1}), the adjacency matrix $A$ is another group of variables in the loss function. We thus can use the gradient information of the adjacency matrix to realize the attack, i.e., lead to an error in the node classification.

Based on the trained GCN model, we further design a target loss function $L_t$ as
\begin{equation}
L_t=-\sum_{k=1}^{|F|}Y_{tk}\ln(Y'_{tk}(A)),
\label{equ:loss2}
\end{equation}
which represents the difference between the predicted label and the real one of the target node $v_t$. The larger value of this loss function corresponds to the worse prediction result. We then calculate the partial derivatives of the target loss function $L_t$ with respect to the element of adjacency matrix, $A_{ij}$, in the network, and further obtain all the link gradient matrix $g$, represented by
\begin{equation}
g_{ij}=\frac{\partial L_t}{\partial A_{ij}},
\label{equ:gradient}
\end{equation}
with $A_{ij}$ being an element of $A$.

Here, we aim to maximize the target loss function $L_t$. Generally, link changes along the same direction of the gradient can make the target loss function $L_t$ increase fastest locally, resulting in the misclassification of the target node quickly using the trained GCN model. Considering that the adjacent matrix of an undirected network is symmetry, here we symmetrize $g$ to obtain $\hat{g}$.
\begin{equation}
\hat{g}_{ij}=\hat{g}_{ji} = \left\{
\begin{array}{ll}
\frac{g_{ij}+g_{ji}}{2} & i\ne j \\
0 & i = j
\end{array}\right.
\label{equ:gradient1}
\end{equation}
We treat $\hat{g}$ as a link gradient network (LGN), where each pair of nodes could be connected with a positive or negative weight denoting the link gradient, with the following meanings:
\begin{itemize}
\item \textbf{Sign:} A positive/negative link gradient $\hat{g}_{ij}$ indicates that adding/deleting the link between the pair of nodes $(v_i,v_j)$ will increase the target loss function.
\item \textbf{Magnitude:} The larger magnitude of link gradient $\hat{g}_{ij}$ means the added/deleted link between the pair of nodes $(v_i,v_j)$ can influence the classification result of the target node more significantly.
\end{itemize}

\subsubsection{Adversarial network generator}
Based on Eq.~(\ref{equ:gradient}), we propose a model to generate adversarial networks to realize the efficient attack on the original network. In this model, we modify a link during each iteration, and the process lasts for $K$ iterations in total. The $h^{th}$ iteration can be described by the following steps.
\begin{itemize}
\item \textbf{Constructing the LGN:} Based on Eq.~(\ref{equ:gradient}) and Eq.~(\ref{equ:gradient1}), we generate the $(h-1)^{th}$ LGN $\hat{g}^{h-1}$ using the adversarial network adjacency matrix $\hat{A}^{h-1}$, with $\hat{A}^0=A$.
\item \textbf{Selecting the target pair of nodes:}
Based on $\hat{g}^{h-1}$, we select a pair of nodes $(v_i,v_j)$ of the maximum absolute link gradient. Note that, if they have positive/negative gradient and meanwhile are connected/disconnected in the original network, we cannot further add/remove the link between them. Therefore, we just ignore such pairs of nodes in the process.
\item \textbf{Realize the attack:}
We use the selected pair of nodes $(v_i,v_j)$ to attack the $(h-1)^{th}$ adversarial network, and generate the adversarial network $\hat{G}^h$. The adjacency matrix $\hat{A}^h$ of the $h^{th}$ adversarial network is defined as:
\begin{equation}
\hat{A}^h_{ij}=\hat{A}^{h-1}_{ij}+\theta (\hat{g}_{ij}),
\label{equ:iteration}
\end{equation}
where $\hat{A}^h_{ij}$ and $\hat{A}^{h-1}_{ij}$ are the elements of $\hat{A}^h$ and $\hat{A}^{h-1}$, respectively, and $\theta(\hat{g}_{ij})$ represents the sign of gradient of the selected pair of nodes $(v_i,v_j)$.

\end{itemize}
The pseudo-code for the adversarial network generator is given in Algorithm~1.
\begin{figure*}[!htbp]
  \centering
  \includegraphics[width=\linewidth]{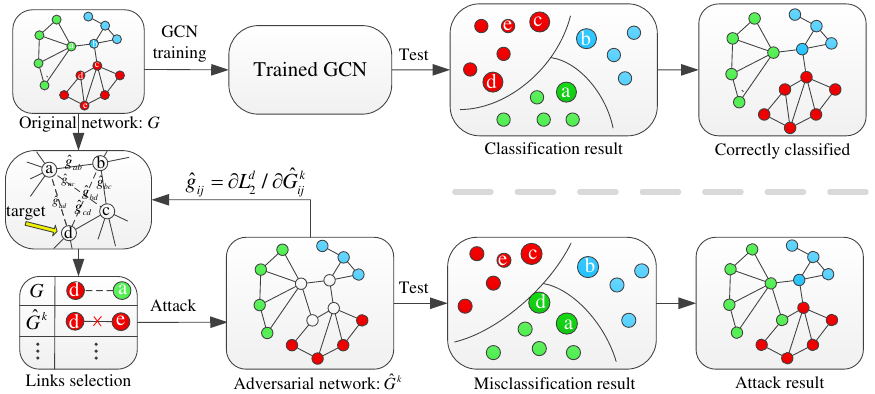}
  \caption{Adversarial network attack generator via GCN.}
  \label{fig:GCNAttack}
\end{figure*}
\begin{algorithm}[h]
\caption{Adversarial network generator via GCN}
\LinesNumbered
\KwIn{Original network $G$, number of iterations $K$.}
\KwOut{The adversarial network $\hat{G}$.}
Train the GCN model on original network $G$ to obtain $\hat{g}^0$ via Eq.~(\ref{equ:gradient}) and Eq.~(\ref{equ:gradient1})\;
Initialize the adjacency matrix of the adversarial network by $\hat{A}^0=A$\;
\For{h = 1 to $K$}{
	Construct $\hat{g}^{h-1}$ based on the $\hat{A}^{h-1}$\;
    Select the pair of nodes $(v_i,v_j)$ of the maximum absolute link gradient in $\hat{g}^{h-1}$\;
    Update the adjacency matrix $\hat{A}^h$ by
    $\hat{A}^h_{ij}=\hat{A}^{h-1}_{ij}+\theta (\hat{g}_{ij})$\;
}
\Return The adversarial network $\hat{G}$, with the adjacency matrix of adversarial network $\hat{A}^K$.\
\end{algorithm}

\subsection{White-Box Adversarial Attack}
Here, we perform the white-box adversarial attack, i.e., using the adversarial network to attack the GCN model. Specifically, the proposed adversarial network generator generates an adversarial network with tiny perturbations, and the re-trained GCN model with the generated adversarial network fails to classify the target nodes correctly. In particular, we will consider direct, indirect, and unlimited attacks, respectively, described as follows.
\begin{itemize}
\item \textbf{Direct attack:} Typically, individuals may have very limited knowledge of the social ties beyond their friends, but rather, they can easily manage their immediate neighborhoods. In order to simulate such cases, we only consider to attack the links around the target nodes, i.e., remove existent links of the target nodes or add new ones to them.
\item \textbf{Indirect attack:} When selecting the links to be modified in the adversarial network attack generator, we find some indirect links (i.e., they are not immediately connected to the target nodes) may also have an impact on the classification result of the target nodes. In certain situations, such indirect links might be preferred to be changed by the network manager in order to make the attack more concealed.
\item \textbf{Unlimited attack:} In this case, we don't limit the attack to the direct or indirect links, i.e., we can remove or add a link between any pair of nodes, in order to seek the maximum attack effect.
\end{itemize}

\subsection{Black-Box Adversarial Attack}
We also perform the adversarial attack generated by GCN to attack other embedding methods, to validate that such attack is quite universal. Since many embedding methods with any downstream classification or regression algorithms have similar decision boundaries, we believe the GCN-based adversarial attack can also be effective on many other embedding methods.

The strong transferability of such adversarial attacks may bring the security concern for network embedding applications, since malicious examples may be easily crafted even when the target network embedding method is unknown in advance. Moreover, in network theory, hub nodes and bridge nodes play important roles in many network dynamics and algorithms. Therefore, we also perform the adversarial attack on such special nodes to see the anti-attack ability of FGA on the critical part of the network.



\section{Experiments\label{Exp}}
In order to testify the effectiveness of our FGA method, we compare it with some baseline attack methods by performing a number of experiments, including uniform attack, hub-node attack, bridge-node attack and community deception. Our experimental environment consists of i7-7700K 3.5GHzx8 (CPU), TITAN Xp 12GiB (GPU), 16GBx4 memory (DDR4) and Ubuntu 16.04 (OS).


\subsection{Datasets}
In the task of node classification, each node in a network is assigned a label, and the following three networks are used. Their basic statistics are summarized in TABLE~\ref{data sets}.
\begin{itemize}
\item \textbf{Pol.Blogs:} The Pol.Blogs dataset is compiled by Adamic and Glance~\cite{adamic2005political}. This dataset is about political leaning collected from blog directories. The blogs are divided into two classes. The links between blogs were automatically extracted from the front pages of the blogs. It contains 1,490 blogs and 19,090 links in total.
\item \textbf{Cora:} This dataset contains a number of machine-learning papers of seven classes~\cite{Mccallum2000Automating}. The links between papers represent the citation relationships. It contains 2,708 papers and 5,429 links in total.
\item \textbf{Citeseer:} This dataset is also a paper citation network with the papers divided into six classes~\cite{Mccallum2000Automating}. It contains 3,312 papers and 4,732 citation links in total.
\end{itemize}

\begin{table}[!htbp]
\centering
\caption{The basic statistics of the three network datasets.}
\label{data sets}
\begin{tabular}{c|ccc}
\hline
\hline
Dataset & \#Nodes & \#Links & \#Classes \\ \hline
Pol.Blogs &1,490  & 19,090 & 2  \\
Cora      & 2,708 & 5,429 & 7      \\
Citeseer  & 3,312 & 4,732 & 6      \\\hline \hline
\end{tabular}
\end{table}

\begin{table*}[!t]
\centering
\caption{The attack effects, in terms of ASR and AML, obtained by different attack methods on varous network embedding methods and multiple datasets. Here, ASR is obtained by changing 20 links.}
\label{one-target attack}
\resizebox{\linewidth}{!}{
\begin{tabular}{c|c|ccc|ccc|ccc|ccc}
\hline
\hline
\multirow{3}{*}{Dataset} & \multirow{3}{*}{NEM} & \multicolumn{6}{c|}{ASR (\%)}        & \multicolumn{6}{c}{AML}          \\ \cline{3-14}
                           &                        & \multicolumn{3}{c|}{FGA} & \multicolumn{3}{c|}{Baseline} & \multicolumn{3}{c|}{FGA} & \multicolumn{3}{c}{Baseline} \\ \cline{3-14}
                           &                        & Unlimited     & Direct     & Indirect     & NETTACK    & DICE    & RA    & Unlimited      & Direct     & Indirect    & NETTACK     & DICE    & RA    \\ \hline

\multirow{7}{*}{Pol.Blogs} & GCN                    & \textbf{87.87}  & 85.74  &25.53    &82.97      & 50.27 & 0.00       & \textbf{8.42}   & 8.82    & 17.61   & 11.89  & 11.85   & 20.00    \\
                           & GraRep                 & \textbf{83.88}  & 81.66  & 4.25    &79.91      & 61.06 & 0.00       & \textbf{9.58}   & 10.42   & 19.36   & 10.48  & 14.22   & 20.00    \\
                           & DeepWalk               & \textbf{84.26}  & 81.66  & 6.25    &75.41      & 64.52 & 0.00       & \textbf{9.84}   & 10.93   & 19.20   & 10.06  & 12.35   & 20.00    \\
                           & node2vec               & \textbf{84.34}  & 81.83  & 0.00    &78.32      & 67.89 & 0.00       & \textbf{9.72}   & 10.16   & 20.00   & 10.58  & 14.86   & 20.00    \\
                           & LINE                   & \textbf{85.25}  & 82.03  & 0.00    &76.35      & 66.74 & 0.00       & \textbf{9.90}   & 11.01   & 20.00   & 10.26  & 12.82   & 20.00    \\
                           & GraphGAN               & \textbf{81.21}  & 80.24  & 0.00    &72.26      & 64.58 & 0.00       & \textbf{9.41}   & 11.02   & 20.00   & 11.08  & 12.26   & 20.00    \\ \cline{2-14}
                           & Average                & \textbf{84.47}  & 82.19  & 6.01    &77.54      & 62.51 & 0.00       & \textbf{9.48}   & 10.39   & 19.36   & 10.73  & 13.06   & 20.00    \\ \hline
\multirow{7}{*}{Cora}      & GCN                    & \textbf{100}    & 100    & 88.28   &92.87      & 54.95 & 6.31       & \textbf{2.54}   & 3.21    & 6.77    & 6.09   & 9.13    & 16.99    \\
                           & GraRep                 & \textbf{100}    & 100    & 84.47   &97.22      & 89.09 & 9.43       & \textbf{5.56}   & 5.57    & 9.41    & 5.94   & 7.37    & 18.43    \\
                           & DeepWalk               & \textbf{100}    & 97.22  & 81.55   &94.06      & 93.52 & 12.16      & \textbf{5.61}   & 6.27    & 10.59   & 7.24   & 7.20    & 17.69    \\
                           & node2vec               & \textbf{100}    & 100    & 84.00   &97.29      & 89.09 & 9.43       & 5.66   & \textbf{5.58}    & 9.52    & 6.75   & 7.37    & 18.43    \\
                           & LINE                   & \textbf{100}    & 96.04  & 84.47   &96.34      & 88.99 & 13.08      & \textbf{5.64}   & 6.36    & 9.41    & 7.02   & 7.66    & 18.07    \\
                           & GraphGAN               & \textbf{100}    & 96.00  & 84.62   &92.26      & 84.55 & 8.49       & \textbf{5.65}   & 6.40    & 11.02   & 8.82   & 7.96    & 18.60    \\ \cline{2-14}
                           & Average                & \textbf{100}    & 98.21  & 84.57   &95.01      & 83.37 & 9.82       & \textbf{5.11}   & 5.57    & 9.62    & 6.98   & 7.78    & 18.04    \\ \hline
\multirow{7}{*}{Citeseer}  & GCN                    & \textbf{100}    & 100    & 91.36   &87.50      & 70.37 & 2.47       & \textbf{3.52}   & 3.88    & 7.69    & 6.88   & 9.87    & 19.36    \\
                           & GraRep                 & \textbf{100}    & 98.39  & 98.41   &94.28      & 93.22 & 32.26      & \textbf{5.32}   & 6.23    & 8.25    & 6.51   & 7.56    & 15.09    \\
                           & DeepWalk               & \textbf{100}    & 100    & 100     &96.96      & 93.44 & 36.51      & \textbf{5.68}   & 6.06    & 7.76    & 7.06   & 7.08    & 14.41    \\
                           & node2vec               & \textbf{100}    & 100    & 98.21   &93.93      & 91.38 & 34.43      & \textbf{5.62}   & 6.50    & 7.75    & 6.34   & 7.13    & 14.87    \\
                           & LINE                   & \textbf{100}    & 100    & 98.36   &95.82      & 96.72 & 32.26      & \textbf{5.88}   & 6.25    & 7.56    & 6.02   & 7.21    & 15.27    \\
                           & GraphGAN               & \textbf{100}    & 97.89  & 93.15   &92.06      & 88.24 & 20.00      & \textbf{5.91}   & 6.67    & 8.18    & 7.42   & 8.26    & 16.55    \\ \cline{2-14}
                           & Average                & \textbf{100}    & 99.38  & 96.75   &93.43      & 88.90 & 26.32      & \textbf{5.16}   & 5.93    & 7.85    & 6.71   & 7.98    & 15.92    \\ \hline \hline
\end{tabular}
}
\end{table*}

\subsection{Baseline Methods}
We compare our FGA method with three network embedding attack methods. Suppose the target node $v_t$ has $n_t$ links in the original network. These baseline methods are briefly described as follows.
\begin{itemize}
\item \textbf{Random Attack (RA)}. RA randomly disconnects $b$ ($b<K$) links in the original network, while randomly connects $K-b$ pairs of nodes that are originally not connected. This is the simplest attack method.
\item \textbf{Disconnect Internally, Connect Externally (DICE)}~\cite{waniek2018hiding}. DICE first randomly disconnect $b$ links of target node, then randomly connect the target node to $K-b$ nodes of different classes.
\item \textbf{NETTACK}~\cite{DBLP:conf/kdd/ZugnerAG18}. NETTACK generates adversarial network iteratively. In each iteration, it selects candidate links based on the degree distribution; then, it defines a scoring function, meaning the confidence loss of the target node in the trained GCN model when a certain link is changed; after that, it utilizes the scores of the candidate links to update the adversarial network.
\end{itemize}
For RA and DICE, we simply set $b=n_t/2$ if $n_t<K$; and $b=K/2$ otherwise.

In order to validate the transfer ability of our FGA method, besides GCN, we also compare it with the three baseline methods on attacking other network embedding approaches including GraRep, DeepWalk, Node2vec, LINE and GraphGAN.

We randomly choose 20\% nodes in a network as the labeled nodes, which are divided into the  training and validation sets of equal size. The rest 80\% nodes are used for testing. For all the network embedding methods, the vector dimension is set to 128, the number of walks per node is set to 10, the length of walk is set to 80 and the size of context window is set to 10. For LINE method, the negative ratio is set to 5. The feature vectors are inputted into a logistic regression classifier to perform node classification with 9:1 train-test ratio.

\subsection{Results}
Now, let's present the attack results obtained by our FGA method and the baseline methods on GCN and other several network embedding methods. We use the following two metrics to measure the attack effectiveness.


\begin{itemize}
\item \textbf{ASR}: The attack success rate, i.e., the ratio of the successfully attacked embeddings of nodes, i.e., leading the misclassification of the node, versus all target nodes, by changing no more than $\gamma$ links for each target node. For a certain value of $\gamma$, the larger ASR corresponds to the better attack effect. Here, the perturbation size $\gamma$ is varied from 1 to 20.
\item \textbf{AML}: The average number of modified links to successfully attack the embedding of a target node. Here, to avoid changing too many links, we limit each method to modify at most 20 links. In other words, if the embedding of the target node is unable to be successfully attacked by modifying 20 links, we simply set the number as 20. The smaller AML corresponds to the better attack effect.
\end{itemize}

\begin{figure*}[!htbp]
\centering
\includegraphics[width=1\linewidth]{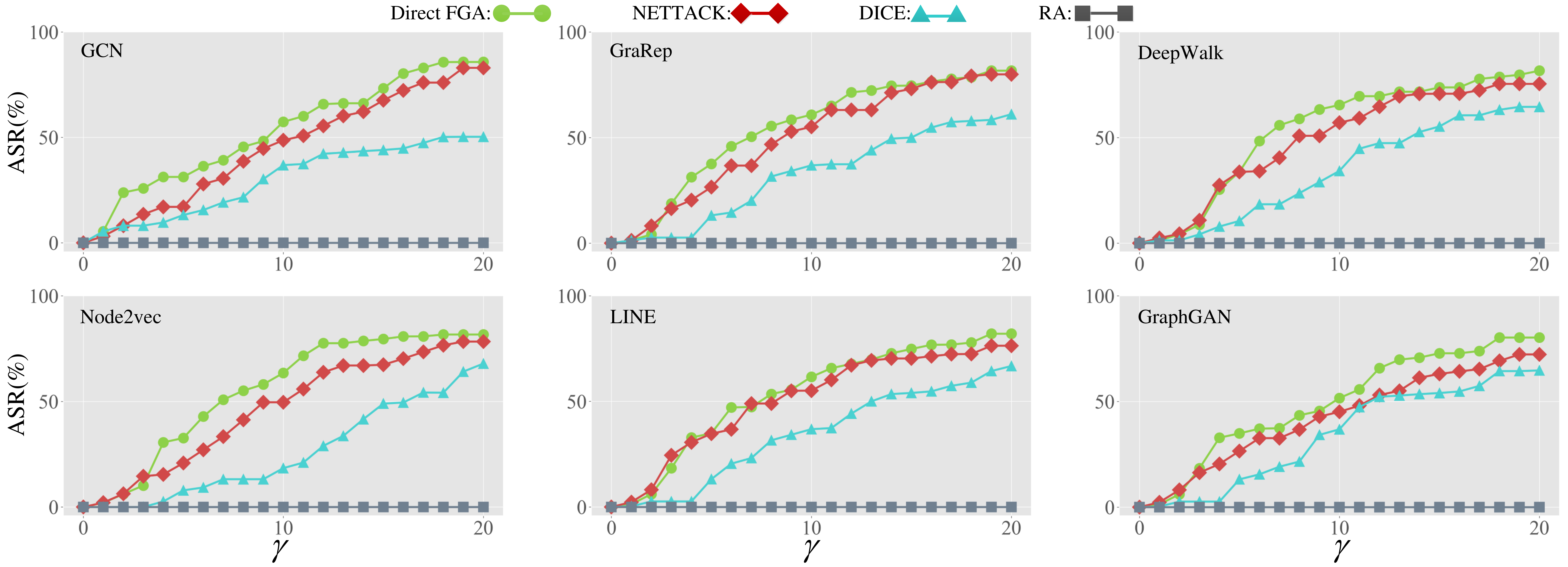}
\caption{ASR of different attack methods as functions of perturbation size $\gamma$ on various network embedding methods for Polblogs dataset.}
\label{Polblogs_AR}
\end{figure*}
\begin{figure*}[!htbp]
\centering
\includegraphics[width=1\linewidth]{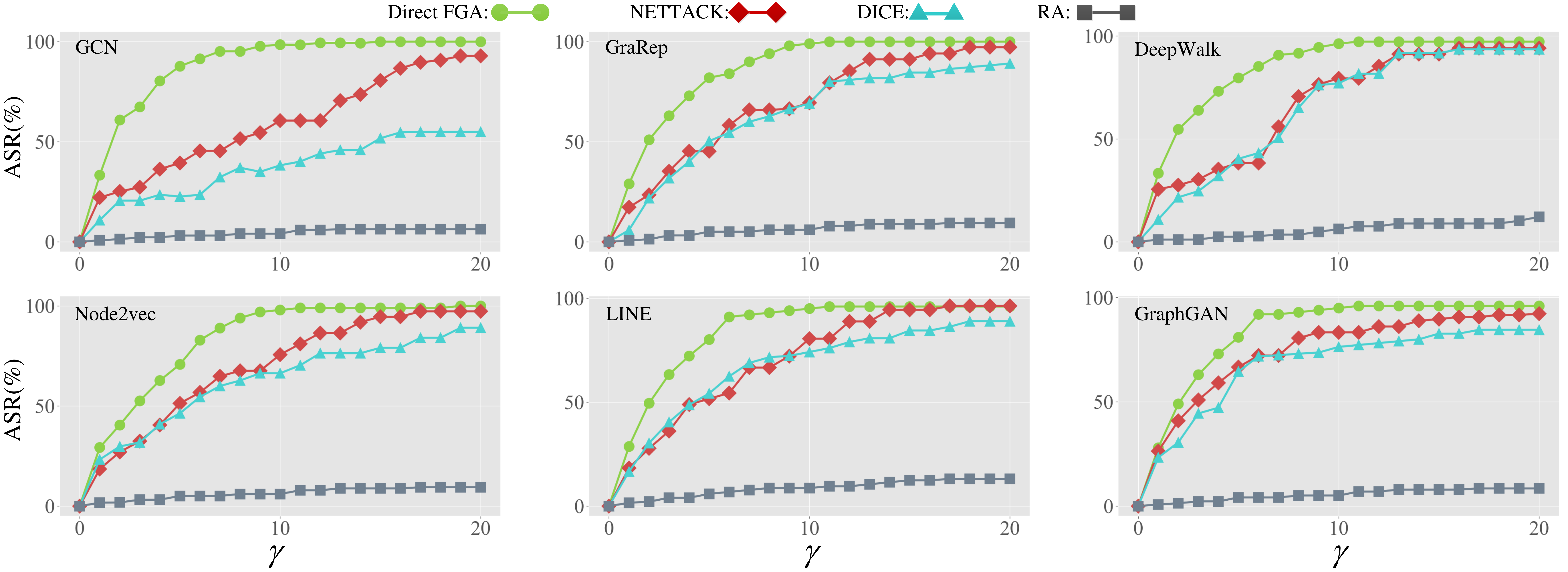}
\caption{ASR of different attack methods as functions of perturbation size $\gamma$ on various network embedding methods for Cora dataset.}
\label{Cora_AR}
\end{figure*}
\begin{figure*}[!htbp]
\centering
\includegraphics[width=1\linewidth]{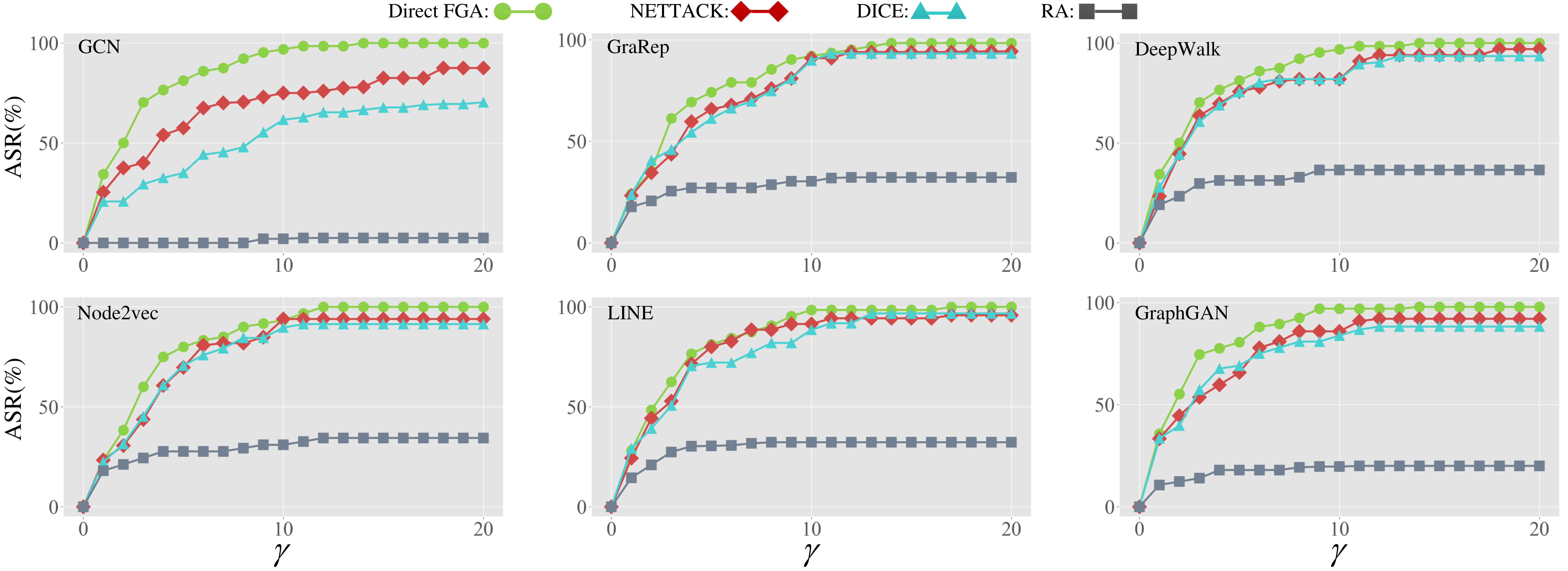}
\caption{ASR of different attack methods as functions of perturbation size $\gamma$ on various network embedding methods for Citeseer dataset.}
\label{Citeseer_AR}
\end{figure*}


\subsubsection{Uniform attack}
In this part, for each network, we random select 20 nodes in each category as the target nodes. The attack results are presented in TABLE~\ref{one-target attack}, where we can see that the unlimited FGA outperforms all the other attack methods in all the cases, in terms of higher ASR and lower AML. Surprisingly, for the datasets of Cora and Citeseer, we can achieve 100\% ASR when 20 links are changed for each target node. While for Pol.Blogs, we can only get 84.47\% ASR on average, which may be because this network is relatively dense, with the average degree close to 25.6. Even though, the unlimited FGA still performs best on attacking any considered network embedding method. It might be argued that, for NETTACK and DICE, only those links around the target nodes are allowed to be changed, i.e., they can be considered as direct attacks. Therefore, it seems to be fairer to compare them with our direct FGA, rather than unlimited FGA. In fact, we can still find that our direct FGA presents better attack effect than NETTACK and DICE in most cases.

In particular, for the datasets of Cora and Citeseer, on average, we only need to change no more than 6 links to successfully attack the embedding of a target node for any network embedding method considered in this paper, by using our unlimited FGA. By comparison, we need to change more links in Pol.Blogs to successfully attack the network embedding methods, due to its denser structure. Overall, although our FGA attack is based on GCN, it has strong trasnferability to attack other network embedding methods, i.e., in TABLE~\ref{one-target attack} we can see that both unlimited and direct FGA achieve higher ASR and lower AML than the three baseline attack methods, while DICE and RA are not designed for a particular network embedding method. On the other hand, white-box attack indeed performs slightly better than black-box attack, e.g., for the datasets of Cora and Citeseer, when unlimited FGA and direct FGA are used to attack GCN, on average, only less than 4 links need to be changed to make the attack successful.

More interestingly, for the datasets of Cora and Citeseer, where the networks are relatively sparse, indirect FGA can also achieve reasonable attack effect, similar to DICE. This indicates that we may also change the links far from the target nodes to realize the attack. In other words, the local structure of these nodes is not necessarily destroyed, making the attack more concealed. However, again it seems that indirect FGA doesn't work on Pol.Blogs, i.e., it only performs better than random attack. This is reasonable, since changing only 20 links far from a target node of mean degree larger than 25 seems not enough to cheat the algorithm to classify it into a wrong group.

Moreover, we also calculate the attack effect, in terms of ASR, obtained by different attack methods, under different perturbation size $\gamma$, varied from 1 to 20. Here, direct FGA is adopted to make fair comparison. The results are shown in Fig.~\ref{Polblogs_AR}-Fig.~\ref{Citeseer_AR}, for different datasets, where we can find that the direct FGA performs best no matter what perturbation size is adopted; NETTACK is better than DICE in most cases; while all of these heuristic and adversarial attacks behave much better than the random attack, indicating that network structure does matter in attacking network embedding methods.

\begin{figure}[!h]
\centering
\includegraphics[width=1\linewidth]{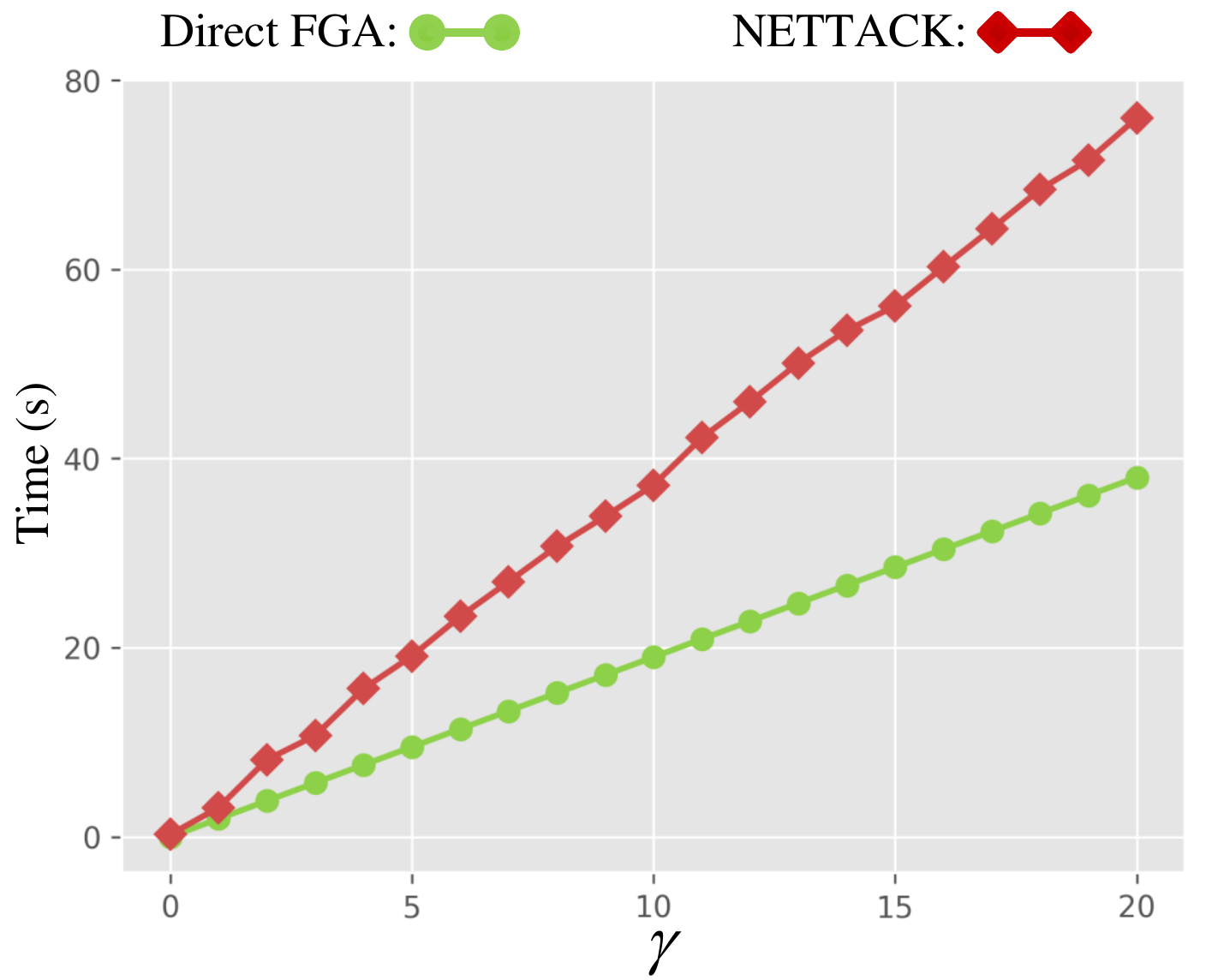}
\caption{The running time of direct FGA and NETTACK.}
\label{time}
\end{figure}

Lastly, we want to compare the time complexity between the direct FGA method and the NETTACK method, as two comparable adversarial network attack methods. Without loss of generality, we record their running time on Cora dataset. In Fig.~\ref{time}, we can see that the running time of both direct FGA and NETTACK increases linearly with the number of modified links, while the running time of NETTACK is significantly more than that of FGA. This is reasonable, because NETTACK needs to select candidate links based on the degree distribution in each iteration, which largely increases the time complexity of this method.

\begin{table*}[!t]
\centering
\caption{The attack effects on hub node, in terms of ASR and AML, obtained by different attack methods on various network embedding methods and multiple datasets. Here, ASR is obtained by changing 20 links.
}
\label{hub}
\resizebox{\linewidth}{!}{
\begin{tabular}{c|c|ccc|ccc|ccc|ccc}
\hline
\hline
\multirow{3}{*}{Dataset} & \multirow{3}{*}{NEM} & \multicolumn{6}{c|}{ASR (\%)}        & \multicolumn{6}{c}{AML}          \\ \cline{3-14}
                           &                        & \multicolumn{3}{c|}{FGA} & \multicolumn{3}{c|}{Baseline} & \multicolumn{3}{c|}{FGA} & \multicolumn{3}{c}{Baseline} \\ \cline{3-14}
                           &         & Unlimited     & Direct & Indirect& NETTACK& DICE & RA     & Unlimited       & Direct  & Indirect& NETTACK& DICE    & RA    \\ \hline

\multirow{7}{*}{Pol.Blogs} & GCN     & \textbf{50.00}&45.00   &5.00    &10.00  &10.00  & 0.00   & \textbf{14.78}   &15.28    &19.70    &18.95   &19.50    & 20.00    \\
                           & GraRep  & \textbf{62.50}&57.50   &2.50    &5.00   &17.50  & 0.00   & \textbf{15.72}   &16.05    &19.55    &19.45   &19.20    & 20.00    \\
                           & DeepWalk& \textbf{53.85}&50.00   &2.50    &12.50  &22.50  & 0.00   & \textbf{16.00}   &16.82    &19.77    &19.20   &19.18    & 20.00    \\
                           & node2vec& 32.50&\textbf{45.00}   &5.00    &12.50  &17.50  & 0.00   & \textbf{17.62}   &17.95    &19.20    &19.20   &19.05    & 20.00    \\
                           & LINE    & \textbf{20.26}&12.21   &5.00    &5.00   &20.00  & 0.00   & \textbf{18.57}   &19.02    &19.31    &19.30   &19.70    & 20.00    \\
                           & GraphGAN& \textbf{25.00}&17.50   &2.50    &2.56   &12.50  & 0.00   & \textbf{18.68}   &18.77    &19.52    &19.67   &19.35    & 20.00    \\ \cline{2-14}
                           & Average & \textbf{40.69}&37.86   &3.75    &7.92   &16.67  & 0.00   & \textbf{16.90}   &17.32    &19.51    &19.30   &19.16    & 20.00  \\\hline
\multirow{7}{*}{Cora}      & GCN     & \textbf{88.90}&87.18   &54.54   &85.71  &48.57  & 4.74   & \textbf{6.48}    &6.62     &14.37    &8.64    & 11.38   & 18.25    \\
                           & GraRep  & \textbf{84.26}&78.69   &48.52   &77.50  &54.86  & 6.74   & \textbf{7.90}    &8.08     &12.71    &9.97    & 11.06   & 18.16    \\
                           & DeepWalk& \textbf{85.63}&80.50   &43.75   &77.50  &60.50  & 10.27  & \textbf{7.57}    &7.73     &13.70    &10.47   & 11.24   & 18.18    \\
                           & node2vec& \textbf{81.82}&80.34   &41.96   &74.38  &55.80  & 7.68   & \textbf{7.22}    &7.52     &13.05    &10.81   & 12.43   & 19.70    \\
                           & LINE    & \textbf{85.25}&83.96   &46.90   &77.10  &57.69  & 9.96   & \textbf{8.01}    &8.25     &13.24    &10.00   & 12.60   & 19.00    \\
                           & GraphGAN& \textbf{84.36}&82.57   &48.74   &73.87  &54.81  & 6.20   & \textbf{8.00}    &8.15     &12.98    &9.52    & 11.91   & 19.87    \\ \cline{2-14}
                           & Average & \textbf{85.04}&82.21   &47.40   &77.68  &55.37  & 7.60   & \textbf{7.53}    &7.73     &13.34    &9.90    & 11.77   & 18.86    \\ \hline
\multirow{7}{*}{Citeseer}  & GCN     & \textbf{100}  &100     &65.71   &97.22  &52.06  & 2.42   & \textbf{5.11}    &5.23     &9.77     &7.89    &16.40    & 19.89    \\
                           & GraRep  & \textbf{88.89}&87.89   &11.11   &87.30  &61.11  & 5.52   & 12.06   &\textbf{11.61}    &18.69    &12.16   &15.06    & 19.06    \\
                           & DeepWalk& \textbf{88.89}&86.89   &8.33    &84.44  &61.11  & 6.34   & \textbf{10.67}   &12.14    &18.89    &13.03   &14.36    & 19.42    \\
                           & node2vec& 86.11&\textbf{88.89}   &13.51   &84.44  &55.56  & 7.62   & 13.08   &\textbf{12.14}    &18.03    &13.19   &14.75    & 19.60    \\
                           & LINE    & \textbf{89.19}&86.11   &13.89   &87.22  &41.67  & 5.66   & \textbf{11.11}   &12.56    &18.39    &13.00   &16.53    & 19.81    \\
                           & GraphGAN& \textbf{89.19}&88.89   &5.56    &87.14  &50.00  & 4.08   & 13.47   &\textbf{11.95}    &19.17    &12.43   &15.17    & 19.78    \\ \cline{2-14}
                           & Average & \textbf{90.38}&90.28   &19.69   &87.96  &53.59  & 5.27   & \textbf{10.92}   &10.94    &17.16    &11.95   &15.38    & 19.59    \\ \hline \hline
\end{tabular}
}
\end{table*}

\begin{table*}[!t]
\centering
\caption{The attack effects on bridge node, in terms of ASR and AML, obtained by different attack methods on various network embedding methods and multiple datasets. Here, ASR is obtained by changing 20 links.}
\label{bridge}
\resizebox{\linewidth}{!}{
\begin{tabular}{c|c|ccc|ccc|ccc|ccc}
\hline
\hline
\multirow{3}{*}{Dataset} & \multirow{3}{*}{NEM} & \multicolumn{6}{c|}{ASR (\%)}        & \multicolumn{6}{c}{AML}          \\ \cline{3-14}
                           &                        & \multicolumn{3}{c|}{FGA} & \multicolumn{3}{c|}{Baseline} & \multicolumn{3}{c|}{FGA} & \multicolumn{3}{c}{Baseline} \\ \cline{3-14}
                           &         & Unlimited     & Direct & Indirect& NETTACK& DICE & RA     & Unlimited       & Direct  & Indirect& NETTACK& DICE    & RA    \\ \hline

\multirow{7}{*}{Pol.Blogs} & GCN     & \textbf{62.16}&54.05   &5.26    &45.95  &8.96   & 0.00    & \textbf{14.65}   &14.95    &19.29    &13.08   &19.35    & 20.00    \\
                           & GraRep  & \textbf{45.95}&44.74   &5.26    &27.03  &13.16  & 0.00    & 17.22   &\textbf{16.47}    &19.42    &18.00   &19.08    & 20.00    \\
                           & DeepWalk& \textbf{36.84}&\textbf{36.84}   &7.89    &28.21  &26.32  & 2.51    & \textbf{17.08}   &17.61    &18.82    &17.31   &17.55    & 19.68    \\
                           & node2vec& 50.00&\textbf{52.63}   &5.26    &28.21  &21.05  & 2.51    & 16.21   &\textbf{16.03}    &19.13    &17.38   &18.39    & 19.80    \\
                           & LINE    & \textbf{45.68}&40.26   &7.89    &31.58  &10.53  & 0.00    & \textbf{16.90}    &17.10    &19.26    &17.21   &19.05    & 20.00    \\
                           & GraphGAN& \textbf{30.77}&25.79   &5.26    &26.32  &10.53  & 0.00    & 17.92   &18.25    &19.37    &\textbf{17.50}   &19.08    & 20.00    \\ \cline{2-14}
                           & Average & \textbf{45.23}&42.39   &6.14    &31.22  &15.09  & 0.84    & \textbf{16.66}   &16.74    &19.22    &16.75   &18.75    & 20.00  \\\hline
\multirow{7}{*}{Cora}      & GCN     & \textbf{92.59}&88.89   &55.56    &86.21  &50.52  &3.06    & \textbf{4.63}    &4.78    &10.93     &7.07   &15.61    &19.90     \\
                           & GraRep  & \textbf{90.91}&90.32   &39.39    &87.10  &72.73  &8.06    & \textbf{7.88}    &8.45    &14.73     &8.97   &10.61    &18.89     \\
                           & DeepWalk& \textbf{93.75}&90.32   &28.12    &82.35  &68.75  &7.79    & \textbf{7.91}    &8.03    &16.03     &8.53   &11.59    &19.31     \\
                           & node2vec& \textbf{93.75}&86.67   &40.00    &84.85  &68.75  &6.90    & \textbf{8.25}    &9.03    &14.46     &8.61   &10.38    &19.56     \\
                           & LINE    & \textbf{93.33}&88.31   &42.56    &85.35  &73.33  &7.02    & \textbf{7.67}    &8.52    &14.02     &9.06   &11.10    &19.00     \\
                           & GraphGAN& \textbf{90.62}&90.00   &33.33    &87.50  &60.00  &5.85    & 8.69    &8.87    &15.97     &\textbf{8.44}   &11.83    &19.64     \\ \cline{2-14}
                           & Average & \textbf{92.49}&89.09   &39.83    &85.56  &65.68  &6.45    & \textbf{7.51}    &7.95    &14.36     &8.45   &11.85    &19.38   \\ \hline
\multirow{7}{*}{Citeseer}  & GCN     & \textbf{96.97}&\textbf{96.97}   &66.67    &82.59  &58.32  &3.25    & \textbf{4.94}    &5.03    &11.61     &7.63   &14.90     &19.53     \\
                           & GraRep  & \textbf{96.77}&96.67   &35.48    &89.67  &70.00  &10.24   & \textbf{8.94}    &9.50    &16.61     &9.67   &12.17    &18.96     \\
                           & DeepWalk& \textbf{93.33}&\textbf{93.33}   &20.69    &86.67  &70.97  &12.06   & \textbf{8.80}    &9.33    &18.76     &9.40   &12.90    &18.65     \\
                           & node2vec& \textbf{93.33}&90.32   &26.67    &86.55  &76.67  &11.64   & 9.63    &\textbf{9.06}    &16.27     &10.21  &10.87    &19.02     \\
                           & LINE    & \textbf{94.44}&91.02   &26.67    &83.33  &74.19  &12.04   & \textbf{8.89}    &9.20    &15.92     &10.87  &11.00    &18.37     \\
                           & GraphGAN& \textbf{93.55}&93.33   &19.35    &82.59  &66.67  &9.50    & 8.97    &\textbf{8.33}    &17.87     &9.22   &12.87    &19.26    \\ \cline{2-14}
                           & Average & \textbf{94.73}&93.61   &33.29    &85.23  &69.47  &9.79    & \textbf{8.36}    &8.41    &16.17     &9.50   &12.45    &18.97     \\ \hline \hline
\end{tabular}
}
\end{table*}

\subsubsection{Hub-node attack}
The nodes of higher centrality are considered as hubs, which play important roles in many network dynamics. Various centrality metrics~\cite{tarkowski2016closeness,jeong2001lethality,crucitti2006centrality} were proposed, and we use the degree centrality~\cite{yustiawan2015degree} here. We then choose 40 hub nodes of largest degree in each network as our target nodes. The attack results on the hub nodes are presented in TABLE~\ref{hub}.



Again, we find that the unlimited FGA still outperforms all the other considered attack methods. The direct FGA follows, which performs better than the NETTACK and DICE in most cases, indicating the effectiveness of our FGA method on disturbing network embeddings of hub nodes in a network. The difference is that, at this time, the indirect FGA loses its effectiveness even in the dense networks of Cora and Citeseer, suggesting that changing the links far away from hub nodes has little influence on these nodes, since they always have a lot of neighbors, making their embeddings much more robust. Overall, by comparing TABLE~\ref{one-target attack} and TABLE~\ref{hub}, we can find that the embeddings of hub nodes are relatively difficult to attack, i.e., the same attack methods obtain smaller ASR and larger AML on the corresponding same network embedding methods and datasets, for the hub nodes than for the normal ones.

\subsubsection{Bridge-node attack}
Since many real-world networks have modular and hierarchical structure, those nodes connecting different communities play key roles to make the whole network connected and further dominate the information spreading on the network. These nodes, namely \emph{bridge nodes}, are thus as important as hub nodes and should be preferentially protected. The bridge nodes are always of higher betweenness centrality~\cite{Shimbel1953Structural,Bavelas1948A} which is defined as the ratio of the shortest paths passing through the target node among all the shortest paths in the network. We thus use this metric to choose 40 nodes of highest betweenness centrality as our bridge nodes, and the attack results on their network embedding are presented in TABLE~\ref{bridge}.

Similarly, the unlimited FGA performs best. By comparison, the direct FGA still performs better than the NETTACK and DICE in most cases, suggesting that our FGA method is also useful in attacking bridge nodes between different communities in a network. Again, the indirect FGA behaves much worse than the direct FGA as well as the NETTACK and DICE, suggesting that changing the links far away can rarely influence these bridge nodes. Overall, by comparing the results in TABLE~\ref{bridge} and those in TABLE~\ref{one-target attack} and TABLE~\ref{hub}, we can find that the embeddings of bridge nodes are relatively difficult to attack than those of normal ones, but are relatively easier to attack than those of hub nodes.



\subsection{Community Deception}
\begin{table*}[!htbp]
\centering
\caption{The attack effects on community detection, in terms of ASR and AML, obtained by different attack methods on various network embedding methods and multiple datasets. Here, ASR is obtained by changing 20 links.}
\label{Community deception}
\resizebox{\linewidth}{!}{
\begin{tabular}{c|c|ccc|ccc|ccc|ccc}
\hline
\hline
\multirow{3}{*}{Data sets} & \multirow{3}{*}{NEM} & \multicolumn{6}{c|}{ASR (\%)}        & \multicolumn{6}{c}{AML}          \\ \cline{3-14}
&                        & \multicolumn{3}{c|}{FGA} & \multicolumn{3}{c|}{Baseline} & \multicolumn{3}{c|}{FGA} & \multicolumn{3}{c}{Baseline} \\ \cline{3-14}
&                        & Unlimited     & Direct     & Indirect   & NETTACK    & DICE    & RA    & Unlimited      & Direct     & Indirect   & NETTACK   & DICE    & RA    \\ \hline
\multirow{6}{*}{PloBook}  & GraRep     & \textbf{96.77}  & 91.67  & 34.92 &89.37   & 74.19 & 3.28      & \textbf{6.77} & 7.97    & 16.33 &10.27    & 10.74   & 19.39    \\
                          & DeepWalk   & \textbf{93.55}  & 88.64  & 22.95 &87.06   & 77.42 & 4.76      & \textbf{7.84} & 8.70    & 18.33 &11.56    & 12.27   & 19.22    \\
                          & node2vec   & \textbf{92.06}  & 90.26  & 19.05 &86.30   & 44.44 & 1.61      & \textbf{8.68} & 9.01    & 18.41 &11.83    & 15.57   & 19.76    \\
                          & LINE       & \textbf{90.48}  & 85.35  & 17.46 &84.12   & 53.97 & 3.17      & \textbf{8.89} & 9.26    & 18.73 &12.69    & 14.25   & 19.52    \\
                          & GraphGAN   & \textbf{93.65}  & 89.21  & 17.46 &81.18   & 53.23 & 3.28      & \textbf{8.32} & 8.34    & 18.63 &12.51    & 14.60   & 19.38    \\
                          \cline{2-14}
                          & Average    & \textbf{93.30}  & 89.03  & 22.37 &85.61   & 60.65 & 3.22      & \textbf{8.10} & 8.66    & 18.09 &11.77    & 13.49   & 19.45    \\ \hline
\multirow{6}{*}{Dolphin}  & GraRep     & \textbf{100}    & 95.35  & 54.55 &21.05   & 59.09 & 4.76      & \textbf{3.71} & 4.59    & 14.32 &18.20    & 13.23   & 19.05    \\
                          & DeepWalk   & \textbf{100}    & 100    & 68.18 &10.00   & 72.73 & 9.09      & \textbf{3.64} & 4.39    & 13.45 &18.15    & 13.18   & 18.45    \\
                          & node2vec   & \textbf{100}    & 100    & 63.64 &20.00   & 57.14 & 9.09      & \textbf{3.41} & 4.25    & 14.00 &18.50    & 13.24   & 18.27    \\
                          & LINE       & \textbf{100}    & 98.69  & 59.09 &10.00   & 68.18 & 9.52      & \textbf{4.00} & 4.40    & 13.73 &19.65    & 12.87   & 18.81    \\
                          & GraphGAN   & \textbf{100}    & 96.57  & 50.00 &5.26    & 76.19 & 10.00     & \textbf{4.00} & 4.61    & 14.50 &19.50    & 11.76   & 18.10    \\
							\cline{2-14}
                          & Average    & \textbf{100}    & 98.12  & 59.09 &13.26   & 66.67 & 8.49      & \textbf{3.75} & 4.45    & 14.00 &18.80    & 12.86   & 18.54    \\ \hline \hline
\end{tabular}}
\end{table*}



\begin{figure*}[!t]
\centering
\includegraphics[width=\linewidth]{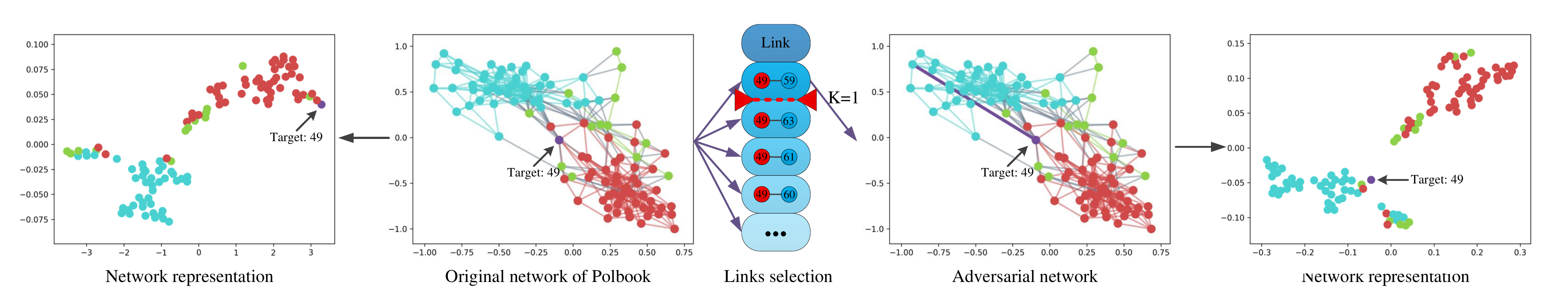}
\caption{The visualization of FGA on network embedding of a random target node in PolBook. The purple node represents the target node and the purple link is selected by our FGA due to its largest gradient. Except for the target node, the nodes of same color belongs to the same community.}
\label{visual: Polbook}
\end{figure*}

\begin{figure*}[!t]
\centering
\includegraphics[width=\linewidth]{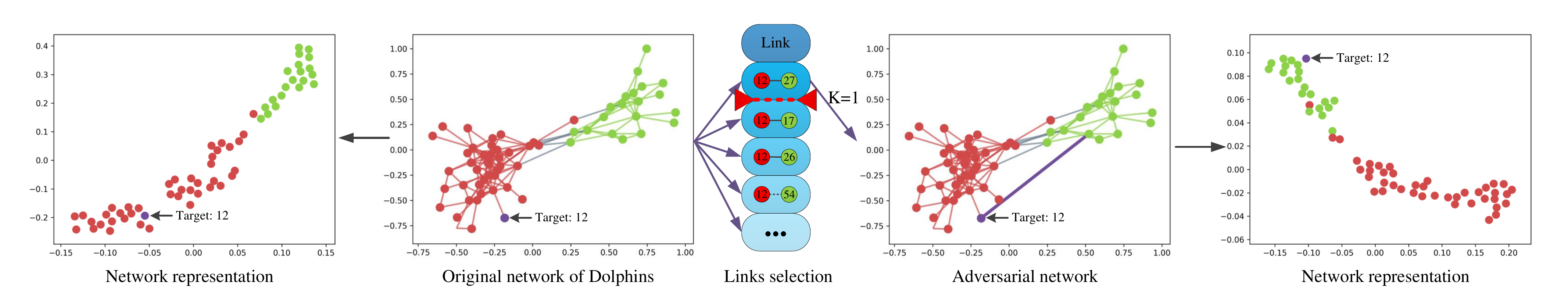}
\caption{The visualization of FGA on network embedding of a random target node in Dolphins. The purple node represents the target node and the purple link is selected by our FGA due to its largest gradient. Except for the target node, the nodes of same color belongs to the same community.}
\label{visual: Dolphins}
\end{figure*}



Network embedding can map a network into a vector space, where each node is represented by a vector of relatively low dimension. As a result, many network algorithms, such as community detection~\cite{cavallari2017learning,zheng2016node,wang2017community}, can be realized in this vector space by using some machine learning methods. Community deception~\cite{nagaraja2010impact,waniek2018hiding,fionda2018community}, on the contrary, is used to protect certain nodes from identifying by community detection methods. Since our FGA methods can be used to disturb network embeddings, we thus think that they can also be used in community deception. In this experiment, we first use GCN to generate the adversarial network; then we use the network embedding methods to generate node embedding vectors, based on which we get the community detection results by adopting the simple K-means method. We use the following two datasets for community detection.

\begin{itemize}
\item \textbf{PloBook:} This network represents co-purchasing of books about US politics sold by the online bookseller~\cite{newman2006modularity}. In this network, nodes represent books about US politics sold by the online bookseller Amazon.com, and two nodes are connected if the corresponding two books were co-purchased by the same buyers. There are 105 books and 441 links in total, and the books belongs to three communities, namely liberal, neutral and conservative.

\item \textbf{Dolphins:} This social network represents the common interactions observed between a group of dolphins in a community living off Doubtful Sound, New Zealand~\cite{lusseau2003bottlenose}. There are 62 dolphins and 159 links in total, and the dolphins are partitioned into two groups by the temporary disappearance of dolphin number.
\end{itemize}

In this experiment, suppose we know the community that each node belongs to in advance. For each network, we randomly select 20\% nodes in each category to train our GCN, with the training and validation sets being of equal size, while the rest 80\% nodes are used for testing. For all the network embedding methods considered here, the dimension of embedding vector is set to 20. Note that, without attack, the nodes in an original network can also be turned into vectors by each network embedding method, and then are grouped into communities by K-means method. There might be some nodes that are wrongly clustered themselves without any attack, and these nodes will not be considered as the target nodes here, i.e., we only attack those nodes that can be corrected clustered initially.

The community deception results are shown in TABLE~\ref{Community deception}, where we can see that these results are consistent with those in node classification, i.e., for each network embedding method on each dataset, the unlimited FGA outperforms all the others, direct FGA behaves better than NETTACK and DICE, while all of them behaves much better than the random attack. These results indicate that our FGA can also be used to attack community detection algorithms by disturbing network embeddings. Note that, we exclude GCN here, since GCN is in nature a supervised learning method, and thus is not suitable for community detection.

In order to make our FGA method easier to understand, we visualize the FGA on network embedding of a random target node in PolBook and Dolphins, as shown in Fig.~\ref{visual: Polbook} and Fig.~\ref{visual: Dolphins}, respectively. Here, we adopt the unlimited FGA and use the low-dimensional network representations learned by DeepWalk as the input to the visualization tool t-SNE~\cite{Maaten2008Visualizing}. We find that the embedding vector of the target node changes a lot even when only one link is changed in each network, indicating the powerful ability of FGA on disturbing network embedding methods.



\section{Conclusion\label{Conclusion}}
In this paper, we propose a framework to generate adversarial networks using GCN, based on which we realize a fast gradient attack (FGA) on network embedding. In this method, we first extract the gradient of pairwise nodes based on the adversarial network, and then select the pair of nodes with maximum absolute link gradient to realize the attack and update the adversarial network, and so forth. This iterative process is terminated when the number of modified links reaches certain predefined value. We conduct numerous experiments, such as uniform attack, hub-node attack and bridge-node attack on six network embedding methods, these embedding vectors are further used to classify nodes in three networks. The results suggest that, in any case, our proposed FGA outperform the other baseline attack methods, achieving the state-of-the-art results. The experiment on community deception also validates the effectiveness of FGA on disturbing network embeddings.


Every coin has two sides, network embedding methods and the corresponding attack methods could be improved iteratively. Therefore, in future, we will also try to propose new network embedding methods that are more robust to the adversarial attacks generated by FGA.

%

\ifCLASSOPTIONcaptionsoff
  \newpage
\fi

\bibliographystyle{IEEEtran}
\bibliography{ref1802}

\end{document}